\documentstyle[12pt,epsf,here,rotate]{article}
\newcommand{\bff}[1]{{\mbox{\boldmath $#1$}}}
\begin{document}
\title{{
TIME DEPENDENT RELATIVISTIC MEAN-FIELD THEORY}}
\author{{ } \\
Dario Vretenar\footnotemark[1] \\
Physik-Department der Technischen Universit\"at M\"unchen}
\date{}
\maketitle
\footnotetext[1]{Alexander von Humboldt Fellow,
on leave of absence from University of Zagreb, Croatia}
\begin{abstract}
\noindent
The relativistic mean-field theory provides a framework in which 
the nuclear many-body problem is described as a self-consistent 
system of nucleons and mesons.
In the mean-field approximation, the self-consistent
time evolution of the nuclear system
describes the dynamics of collective motion: 
double giant resonances, 
nuclear compressibility from monopole resonances,
regular and chaotic dynamics of isoscalar and isovector
collective vibrations.
\end{abstract}
\section{Introduction and outline of the model}

Relativistic mean-field (RMF) models have been successfully applied in
calculations of nuclear matter and properties of finite nuclei throughout
the periodic table. In the self-consistent mean-field
approximation, detailed calculations have been performed for a variety
of nuclear structure phenomena~\cite{Rin.96}. In the present
work we review the applications of RMF to the
dynamics of collective vibrations in spherical nuclei. 
In relativistic quantum hadrodynamics~\cite{SW.86},
the nucleus is described as a system of Dirac nucleons which
interact through the
exchange of virtual mesons and photons.  The Lagrangian
density of the model is
\begin{eqnarray}
{\cal L}&=&\bar\psi\left(i\gamma\cdot\partial-m\right)\psi
~+~\frac{1}{2}(\partial\sigma)^2-U(\sigma )
\nonumber\\
&&-~\frac{1}{4}\Omega_{\mu\nu}\Omega^{\mu\nu}
+\frac{1}{2}m^2_\omega\omega^2
~-~\frac{1}{4}{\vec{\rm R}}_{\mu\nu}{\vec{\rm R}}^{\mu\nu}
+\frac{1}{2}m^2_\rho\vec\rho^{\,2}
~-~\frac{1}{4}{\rm F}_{\mu\nu}{\rm F}^{\mu\nu}
\nonumber\\
&&-~g_\sigma\bar\psi\sigma\psi~-~
g_\omega\bar\psi\gamma\cdot\omega\psi~-~
g_\rho\bar\psi\gamma\cdot\vec\rho\vec\tau\psi~-~
e\bar\psi\gamma\cdot A \frac{(1-\tau_3)}{2}\psi\;.
\label{lagrangian}
\end{eqnarray}
The Dirac spinor $\psi$ denotes the nucleon with mass $m$.
$m_\sigma$, $m_\omega$, and $m_\rho$ are the masses of the
$\sigma$-meson, the $\omega$-meson, and the $\rho$-meson,
and $g_\sigma$, $g_\omega$, and $g_\rho$ are the
corresponding coupling constants for the mesons to the
nucleon. $U(\sigma)$ denotes the nonlinear $\sigma$
self-interaction
\begin{equation}
U(\sigma)~=~\frac{1}{2}m^2_\sigma\sigma^2+\frac{1}{3}g_2\sigma^3+
\frac{1}{4}g_3\sigma^4,
\label{NL}
\end{equation}
and $\Omega^{\mu\nu}$, $\vec R^{\mu\nu}$, and $F^{\mu\nu}$
are field tensors~\cite{SW.86}.

The coupled equations of motion are derived from the Lagrangian
density (\ref{lagrangian}).
The Dirac equation for the nucleons:
\begin{eqnarray}
i\partial_t\psi_i&=&\left[ \bff\alpha
\left(-i\bff\nabla-g_\omega\bff\omega-
g_\rho\vec\tau\vec{\bff\rho}
-e\frac{(1-\tau_3)}{2}{\bff A}\right)
+\beta(m+g_\sigma \sigma)\right.\nonumber\\ 
&&\left. +g_\omega \omega_0+g_\rho\vec\tau\vec\rho_0
+e\frac{(1-\tau_3)}{2} A_0
\right]\psi_i
\label{dirac}
\end{eqnarray}
and the Klein-Gordon equations for the mesons:
\begin{eqnarray}
\left(\partial_t^2-\Delta+m^2_\sigma\right)\sigma&=&
-g_\sigma\rho_s-g_2 \sigma^2-g_3 \sigma^3\\
\left(\partial_t^2-\Delta+m^2_\omega\right)\omega_\mu&=&
~g_\omega j_\mu\\
\left(\partial_t^2-\Delta+m^2_\rho\right)\vec\rho_\mu&=&
~g_\rho \vec j_\mu\\
\left(\partial_t^2-\Delta\right)A_\mu&=&
~e j_\mu^{\rm em}.
\label{KGeq4}
\end{eqnarray}
In the relativistic mean-field approximation, the nucleons
described by single-particle spinors $\psi_i~(i=1,2,...,A)$ are
assumed to form the A-particle Slater determinant $|\Phi\rangle$,
and to move independently in the classical meson fields.
The sources of the fields, i.e.
densities and currents, are calculated in the {\it no-sea}
approximation~\cite{VBR.95}:\ the scalar density:
$\rho_{\rm s}~=~\sum_{i=1}^A \bar\psi_i\psi_i$,
the isoscalar baryon current:
$j^\mu~=~\sum_{i=1}^A \bar\psi_i\gamma^\mu\psi_i$,
the isovector baryon current:
$\vec j^{\,\mu}~=~\sum_{i=1}^A \bar\psi_i\gamma^\mu \vec \tau\psi_i$,
the electromagnetic current for the photon-field:
$j^\mu_{\rm em}~=~\sum_{i=1}^A
\bar\psi_i\gamma^\mu\frac{1-\tau_3}{2}\psi_i$.
The summation is over all occupied states in the
Slater determinant $|\Phi\rangle$. Negative-energy states
do not contribute to the densities in the {\it no-sea}
approximation of the stationary solutions.  
It is assumed that nucleon single-particle
states do not mix isospin. 

The ground state of a nucleus is described by
the stationary self-consistent solution of the
coupled system of equations
(\ref{dirac})--(\ref{KGeq4}),
for a given number of nucleons
and a set of coupling constants and masses.
The solution for the ground state specifies part of the initial
conditions for the time-dependent problem.
The dynamics of nuclear collective motion is analyzed in the framework
of time-dependent relativistic mean-field model,
which represents a relativistic generalization of the
time-dependent Hartree-Fock approach. 
For a given set of initial conditions, i.e. initial values
for the densities and currents, nuclear dynamics is
described by the simultaneous evolution of $A$ single-
particle Dirac spinors in the time-dependent mean fields.
Frequencies of eigenmodes are determined from a Fourier
analysis of dynamical quantities. In this microscopic
model, self-consistent time-dependent 
mean-field calculations are performed for multipole excitations.
An advantage of the time-dependent approach is that no assumption
about the nature of a particular mode of vibrations has to be made.
Retardation effects for the meson fields are not
included in the model, i.e. the time derivatives
$\partial_t^2$ in the equations of motions for the meson
fields are neglected. This is justified by the large masses in the meson
propagators causing a short range of the corresponding
meson exchange forces. 
Negative energy contributions are included implicitly in
the time-dependent calculation, since the Dirac equation is
solved at each step in time for a different basis set~\cite{VBR.95}.
Negative energy components with respect to the original
ground-state basis are taken into account automatically,
even if at each time step the {\it no-sea} approximation is applied.

The description of nuclear
dynamics as a time-dependent initial-value problem is
intrinsically semi-classical, since there is no systematic
procedure to derive the initial conditions that
characterize the motion of a specific mode of the nuclear
system. The theory can be quantized by the
requirement that there exist time-periodic solutions of the
equations of motion, which give integer multiples of
Planck's constant for the classical action along one period
~\cite{RVP.96}.  For giant resonances the time-dependence
of collective dynamical quantities is actually not
periodic, since generally giant resonances are not
stationary states of the mean-field Hamiltonian. The
coupling of the mean-field to the particle continuum allows
for the decay of giant resonances by direct escape of
particles.  In the limit of small amplitude oscillations,
however, the energy obtained from the frequency of the
oscillation coincides with the excitation energy of the
collective state. In Refs.~\cite{VBR.95,RVP.96,Vre.97}
we have shown that the model reproduces 
experimental data on giant resonances in spherical nuclei.

\section{Dynamics of collective vibrations}
\subsection{Double Giant Resonances}
The physics of giant resonances in nuclei has been the 
subject of extensive experimental and theoretical 
studies for many years. However, 
only recently substantial evidence has been reported for two-phonon 
states built with giant resonances. Resonant structures 
that were observed in heavy-ion inelastic scattering,
have been interpreted as possible multiple excitations 
of the giant quadrupole resonance. Double giant dipole 
resonances have also been discovered in the neutron
and in the $\gamma$-spectra of nuclei that have been Coulomb
excited in relativistic heavy-ion collisions. In Ref.~\cite{RVP.96}
we have performed time-dependent relativistic 
mean-field calculations and found evidence for modes 
which can be interpreted as double resonances, and which 
in a quantized theory correspond to two-phonon states.

As an example of double resonances in light nuclei, 
we consider the double isoscalar giant quadrupole 
resonance in $^{40}$Ca. The experimental spectrum exhibits
a prominent structure, centered at $34\pm 2$ MeV excitation
energy, with a width of $9\pm2$ MeV. It is interpreted as
the two-phonon state of the single isoscalar GQR at 17.5 MeV.
Using the NL-SH parameter set for the effective Lagrangian,
we have studied isoscalar quadrupole oscillations in $^{40}$Ca.
The quadrupole mode of oscillations is excited by deforming
the spherical solution for the ground-state.
For a specific initial deformation, we follow the 
time-evolution of the collective variable, the quadrupole
moment 
\begin{equation}
q_{20}(t)=\langle\Phi(t)|\hat Q_{20}|\Phi(t)\rangle=\langle\Phi(t)|
r^2 Y_{20} |\Phi(t)\rangle
\label{qua}
\end{equation}
The time-dependent quadrupole moment shown in Fig. 1
corresponds to an initial axial deformation of the baryon density
$\beta = 0.38$. The resulting Fourier spectrum
displays a strong peak at 18.5 MeV, in reasonable agreement with the
experimental data. 
We find evidence for excitation of a higher mode in the
oscillations of the baryon density.
The wave function of the nuclear system is a Slater
determinant at all times, and therefore can be expanded
in the basis of the ground state $|\Phi_0\rangle$
\begin{eqnarray}
|\Phi(t)\rangle=|\Phi_0\rangle +
\sum_{mi} z_{mi}(t)a^+_m a^{}_{i}~|\Phi_0\rangle+
\sum_{m i m^\prime i^\prime}z_{mi}(t) z_{m^\prime i^\prime}(t)
a^+_ma^+_{m^\prime}a^{}_{i^\prime}a^{}_{i}~|\Phi_0\rangle~+~...
\label{slat}
\end{eqnarray}
If the total wave function contains collective $2p-2h$
components, they will be observed in the Fourier
spectrum of the time-dependent baryon density
$\rho_{B}({\bf r},t)~=~\sum_{i=1}^A
\psi^{+}_{i}({\bf r},t) \psi_{i}({\bf r},t)$.
Because of axial symmetry and the isoscalar nature of the excitation,
it is sufficient to consider oscillations in time of the
baryon density on the positive $z$-axis. In Fig. 2 we
display the Fourier transforms of the time-dependent baryon
density for various values of the coordinate $z$.
Two peaks are clearly observed. The first one at
18.5 MeV corresponds to the isoscalar quadrupole resonance. It
gradually increases from the center toward the surface of the
nucleus. If we plot the values of the Fourier transforms at
18.5 MeV as a function of $z$, the resulting curve corresponds to the
transition density. The transition density for the first peak
is typical for isoscalar quadrupole resonances. The second
peak is at 37 MeV, twice the energy of the GQR. It has
a maximum in the center of the nucleus, at first decreases with
$z$, but then appears again on the surface. 
Compared to the GQR transition density, the curve for the 37 MeV
peak displays an additional node.

Periodic solutions of the time-dependent 
Dirac equations can be used to construct the energies of the
many-body system. The energy spectrum can be obtained from a
semi-classical quantization procedure. One finds
periodic solutions such that the mean-field action along
a periodic orbit
\begin{equation}
I~=~\sum_i \int_{t_0}^{t_0 + T} dt \left [
<\psi_i (t)| i\hbar {\partial \over {\partial t}}| \psi_i (t)>
~-~e_i\right ]
\label{action}
\end{equation}
is equal to an integer multiple of the Planck constant
$I~=~n h$. In Eq. (\ref{action}) T is the period of oscillations,
$|\psi_i (t)>$ denotes time-dependent single-nucleon Dirac
spinors, $e_i$ are the corresponding single-nucleon energies
in the unperturbed ground-state, and the summation runs over the
occupied states. Because of the coupling to the continuum in the
mean-field description, giant resonances are not stationary
states of the Hamiltonian. Consequently, a non-periodic dependence
on time is obtained for dynamical quantities. If the damping is very
strong, the giant resonance is not periodic even on the average,
and the quantization condition cannot be applied. However,
if the motion is nearly periodic, i.e. the damping is relatively
weak, the quantization procedure can still be used to calculate
the energies, and the effect of damping can be taken into
account approximately.

In Fig. 3 we display the mean-field action
as function of the excitation energy of the nucleons
and of the initial deformation $\beta$. For the values of
$\beta$ indicated by dots we
have integrated the coupled system of Dirac and
Klein-Gordon equations and calculated the action
integral. The mean-field action 
is a quadratic function of the initial deformation $\beta$,
and an almost perfect linear function of the excitation
energy. Only above 60 MeV a slight deviation from a pure
linear dependence is observed. The values of $<E^*>$ for
which the action is an integer multiple of the Planck
constant are: $I=1~h$ for 18.5 MeV, and $I=2~h$ for 37.1 MeV.
A double giant dipole resonance has
been observed in relativistic Coulomb excitation of
$^{208}$Pb. The single GDR 
is found at $13.3\pm 0.1$ MeV with
a width of $4.1\pm 0.1$ MeV. The sum energy of coincident
photon pairs displays a broad structure at $25.6\pm 0.9$
MeV with a width of $5.8\pm 1.1$ MeV. It is
interpreted as the double GDR. In order to excite
isovector dipole motion we define the initial conditions:
at $t=0$
(in the center of mass system) all protons start moving in the $+z$
direction with velocity $v_\pi$, and all neutrons start
moving in the $-z$ direction with velocity $v_\nu~=~ {Z\over N} v_\pi$.
For the NL-SH parameter set, the Fourier spectrum of
the time-dependent dipole moment displays a strong peak at 12.9 MeV
excitation energy, in good agreement with experimental data.
We have found that the mean-field action is an integer multiple
of the Planck constant: $I=1~h$ for  $<E^* >=12.9$ MeV,
and $I=2~h$ for $<E^* >=25.9$ MeV. Therefore,
the energy of the one-phonon state, calculated from the
mean-field action, coincides with the resonant energy
of the mean peak in the Fourier spectrum, and 
the two-phonon state at
25.9 MeV is in excellent agreement with the experimental
value for the excitation energy of the double GDR. 
\subsection{Monopole Giant Resonances and nuclear compressibility}
The study of isoscalar monopole resonances in nuclei
provides an important source of information on the nuclear
matter compression modulus $K_{\rm nm}$. This quantity is
crucial in the description of properties of
nuclei, supernovae explosions, neutron stars, and heavy-ion
collisions. In principle the value of $K_{\rm nm}$ can be
extracted from experimental energies of isoscalar
monopole vibrations in nuclei (giant monopole resonances GMR). 
However, the complete experimental data set on
isoscalar GMR does not limit
the range of $K_{\rm nm}$ to better than $200 - 350$ MeV.
Microscopic calculations of GMR excitation energies might
provide a more reliable approach to the determination of
the nuclear matter compression modulus.
Modern non-relativistic
Hartree-Fock plus RPA calculations, using both Skyrme and
Gogny effective interactions, indicate that the value of
$K_{\rm nm}$ should be in the range 210-220 MeV.
In relativistic calculations on the other hand, both
time-dependent and constrained RMF results indicate that
empirical GMR energies are best reproduced by an effective
force with $K_{\rm nm}\approx 250 - 270$ MeV.\\
In Ref.~\cite{Vre.97} we have performed time-dependent 
and constrained RMF calculations for monopole giant 
resonances of a number of spherical closed shell nuclei,
from $^{16}$O to the heavy nucleus $^{208}$Pb. For the effective
Lagrangian we have used six parameter sets, which differ
mostly by their prediction for $K_{\rm nm}$, but otherwise
reproduce reasonably well experimental data on nuclear properties.
The idea is to restrict the possible
values of the nuclear matter compression modulus, on the
basis of the excitation energies of giant monopole states
calculated with different effective interactions. In
addition to the four non-linear sets NL1, NL3, NL-SH and
NL2, we have also included two older linear parametrizations, 
HS and L1.
The sets NL1, NL-SH and NL2 have been extensively
used in the description of properties of finite nuclei
\cite{Rin.96}.  In order to bridge the gap between
NL1 ($K_{\rm nm} = 211.7$ MeV), and NL-SH ($K_{\rm nm} =
355.0$ MeV), we have also included a new effective
interaction NL3 ~\cite{LKR.97} ($K_{\rm nm} = 271.8$ MeV).
This new parameter set provides an excellent
description not only for the properties of stable nuclei,
but also for those far from the valley of beta stability.
From the energy spectra and transition densities calculated with
these effective forces, it has been possible to study the
connection between the incompressibility of nuclear matter
and the breathing mode energy of spherical nuclei.
For the isoscalar mode we have found an almost linear
relation between the excitation energy of the monopole
resonance and the nuclear matter compression modulus. 
For the determination of $K_{\rm nm}$
especially relevant are microscopic calculations of 
GMR excitation energies in heavy nuclei. 

The results of
TD RMF calculations for $^{208}$Pb are displayed in Fig. 4:
time-dependent monopole moments 
$\langle r^2(t)\rangle ~=~\frac {1}{A}\langle \Phi(t) |r^2 |\Phi(t)\rangle$
and the corresponding Fourier power spectra for the nonlinear
effective interactions.  As one would expect for a heavy
nucleus, there is very little spectral fragmentation and a
single mode dominates, at least for NL1 and NL3. The
experimental excitation energy $13.7\pm 0.3$ MeV is very
close to the frequency of oscillations obtained with the
NL3 parameter set: 14.1 MeV.  The calculated excitation
energy for the NL1 parameter set ($K_{\rm nm} =211.7$ MeV),
is approximately 1 MeV lower than the average experimental
value. For the linear effective forces HS and L1  the
oscillations are more anharmonic, and the monopole strength
is located well above the experimental GMR energy.
The effective interactions NL1 and NL3 seem to produce GMR
excitation energies which are quite close to the
experimental values. For these two parameter sets
we have calculated the isoscalar giant monopole
resonances in a number of doubly closed-shell nuclei:
$^{40}$Ca, $^{56}$Ni, $^{100,114,132}$Sn, $^{90,122}$Zr, 
$^{146}$Gd. The results are shown in Fig. 5.
The energies of giant monopole states are determined from
the Fourier spectra of the time-dependent monopole moments,
and are displayed as function of the mass number. 
The NL1 excitation energies are systematically lower,
but otherwise the two effective interactions produce
very similar dependence on the mass number. The empirical
curve $E_x \approx 80~A^{-1/3}$ MeV is also included in the
figure, and it follows very closely the excitation energies
calculated with the NL3 parameter set. 
Similar results are obtained from constrained RMF calculations.
Both methods indicate that, in the framework of
relativistic mean field theory, the
nuclear matter compression modulus $K_{\rm nm} \approx 250
- 270$ MeV is in reasonable agreement with the available
data on spherical nuclei. This value is approximately 20\%
larger than the values deduced from recent
non-relativistic density dependent Hartree-Fock
calculations with Skyrme or Gogny forces
\subsection{Regular and chaotic dynamics of collective vibrations}

The atomic nucleus has been used as a laboratory,
both experimentally and theoretically, for investigating
the transition from order to chaos in quantum dynamical systems.
Most of these studies have concentrated on two
major aspects: (i) generic signatures of chaos in
local fluctuations and correlations of nuclear
level distributions, and (ii) chaos in microscopic and
collective dynamics of realistic many-body systems. 
Regular and chaotic dynamics in
giant nuclear oscillations has been the subject of
a number of studies. What has emerged as a very
interesting result is that an undamped collective mode
may coexists with chaotic single-particle motion.
It appears that the slowly vibrating self-consistent
mean field created by the same nucleons averages
out the random components in their motion.
In all investigations the motion of
only one type of particles has been considered.
That is, only the dynamics of isoscalar
collective modes.

We have studied the difference in the
dynamics of isoscalar and isovector collective modes.
In particular, we consider isoscalar and isovector
monopole oscillations in spherical nuclei, but
analogous considerations apply to higher multipolarities.
In Fig. 6 results are shown of time-dependent
relativistic mean-field calculations for isoscalar
and isovector oscillations in $^{208}$Pb.
The experimental isoscalar GMR energy in $^{208}$Pb is
$13.7\pm 0.3$ MeV, and the excitation energy of the
isovector mode is $26\pm 3$ MeV. Calculations have been 
performed for the NL1 effective interaction.
In the isoscalar case both proton and
neutron densities are radially expanded, while for the
isovector mode the proton density is initially compressed
by the same amount. Therefore, in both cases we follow
the time evolution of the same system, just the initial
conditions are different.
In Fig. 6a we plot the time history of the isoscalar
monopole moment $\langle r^2(t)\rangle$, and in Fig. 6b
the corresponding isovector moment
$< r^2_{\rm p}(t) > - < r^2_{\rm n} (t)>$ is displayed. The
isoscalar mode displays regular undamped oscillations,
while for the isovector mode we observe strongly
damped anharmonic oscillations.
On the right hand panels we
plot the corresponding Fourier power spectra.
The Fourier spectrum of the isovector mode is strongly fragmented.
However, the main peaks are found in the energy region $25 - 30$ MeV,
in agreement with the experimental data.
For the isoscalar mode, the time history of the monopole moment
and the Fourier spectrum show that the oscillations of the collective
coordinate are regular. On the other hand, the appearance of a
broad spectrum of frequencies seems
to indicate that the isovector oscillations are chaotic.

A diagnosis of chaotic vibrations would imply that one has
a clear definition of such motion. For a quantum system,
however, the concept of chaos, especially in time-dependent
problems, is not well defined. And although our description
of nuclear vibrations is semi-classical, quantum effects
like the Pauli principle are present in the initial
conditions and during the dynamical evolution. There
exists a number of tests that can help to
identify chaotic oscillations in physical systems,
and some of them can be applied in the present consideration.
In Figs. 7 - 9 we display some additional qualitative measures
which can be used to characterize the response of our
nonlinear system. In Fig. 7 we have constructed the two-dimensional
time-delayed pseudo-phase space for the isoscalar (a) and
isovector (b) oscillations, shown in Fig. 6. Since information
is available on the time evolution of just one variable, the
collective coordinate, one plots the signal versus itself, but
delayed or advanced by a fixed time constant $[<r^2(t)>,
<r^2(t+\tau)>]$. The phase space trajectories for
the isoscalar mode are closed ellipses, indicating regular oscillations.
For the isovector oscillations on the other hand, the trajectories
are completely chaotic. The strong damping results from
one-body processes: (i) escape of nucleons into the continuum
states and (ii) collisions of the
nucleons with the moving wall of the nuclear potential
generated by the self-consistent mean fields.
In Fig. 8 we display the corresponding Poincar\' e sections
constructed from three-dimensional time-delayed pseudo-phase space.
The Poincar\' e map for the isoscalar mode consists of two sets
of closely located points, and therefore confirms regular oscillations.
For the isovector oscillations the Poincar\' e map appears as
a cloud of unorganized points in the phase plane. Such a map
indicates stochastic motion. 
Another measure that is related to
the Fourier transform is the autocorrelation function
\begin{equation}
A(\tau) = \lim_{T\to \infty} \int\limits_{0}^{T}
	   <r^2(t)> <r^2(t+\tau)>\,dt
\end{equation}
When the signal is chaotic, information about its past origins
is lost. This means that $A(\tau) \to 0$ as $\tau \to \infty$,
or the signal is only correlated with its recent past. 
The autocorrelation functions for isoscalar and isovector oscillations
are shown in Fig. 9. 
For the isovector mode $A(\tau)$ displays a rapid decrease, and the
envelope appears as an
irregular waveform. 
The oscillations of the collective
coordinate can be characterized as regular for the
isoscalar mode, and they become chaotic when
initial conditions correspond to the isovector mode.
In Ref.~\cite{Vre.97a} we have shown how a regular collective mode
can coexist with chaotic single-particle dynamics,
a result which confirms the conclusions of a number of studies.
However, we have also shown that this is the case only
for isoscalar modes, that is, only if one considers
the motion of a single type of particles. When
protons and neutrons move out of phase, as it happens
for isovector modes, the resulting dynamics of the collective
coordinate exhibit chaotic behavior. This is
explained by the fact that protons and neutrons effectively
move in two self-consistent potentials that oscillate out
of phase. For example, when neutrons move inward, they
scatter on the potential wall with positive curvature
that is created by protons moving outward. This will
lead to pseudo-random motion of the nucleons and
dissipation of collective oscillations.

\newpage
\leftline{\bf Figure Captions}
\begin{description} 
\item{\small {\bf FIG. 1.}}
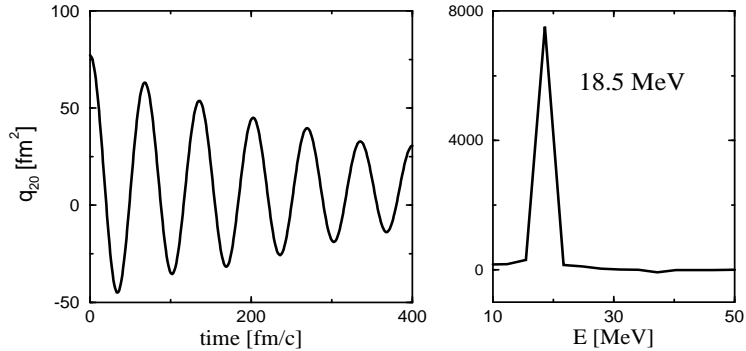
Time-dependent quadrupole moment
and the corresponding Fourier spectrum for $^{40}$Ca.
\item{\small {\bf FIG. 2.}}
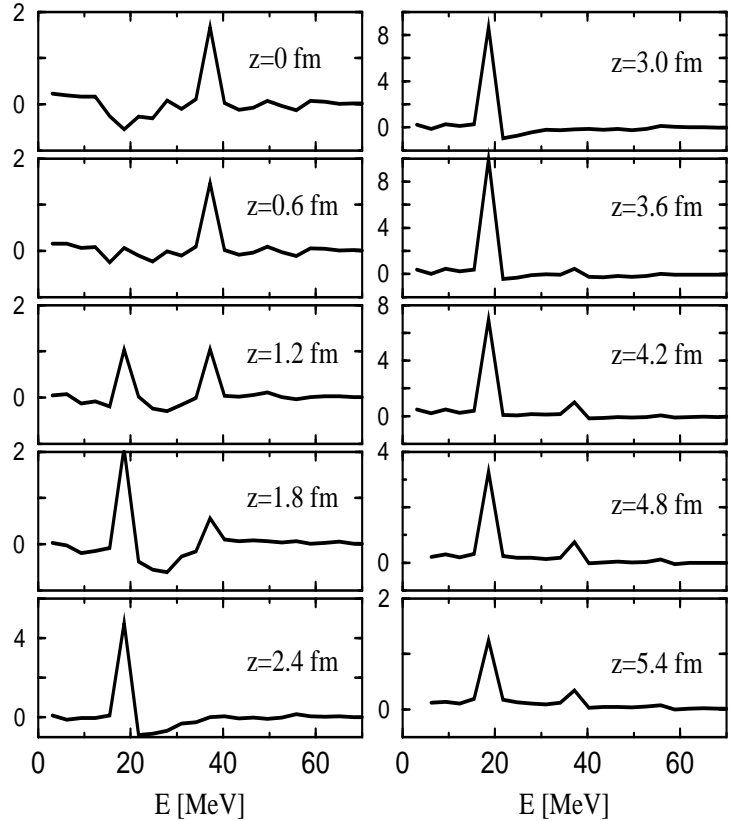
Fourier spectra of the time-dependent
baryon density for various values of the coordinate $z$, on
the axis along which the initial densities are deformed.
\item{\small {\bf FIG. 3.}}
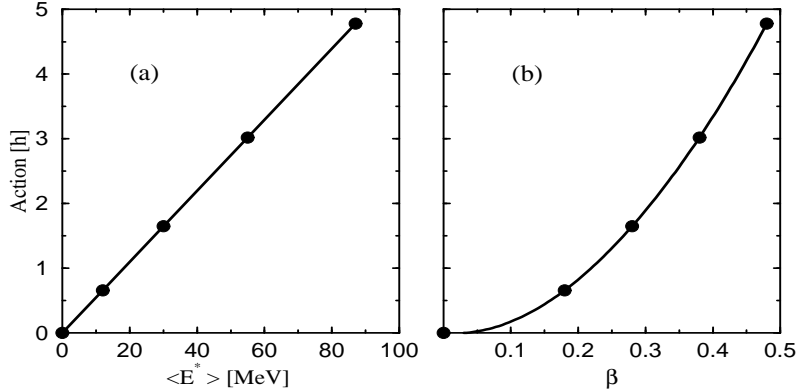
Mean-field action as
function of the excitation energy of the nucleons
$E^*=<\Phi(t)|H_D(t)|\Phi(t) > - <\Phi_{GS} |H_D(0)|\Phi_{GS}>$ (a),
and of the initial deformation $\beta$ (b). 
\item{\small {\bf FIG. 4.}}
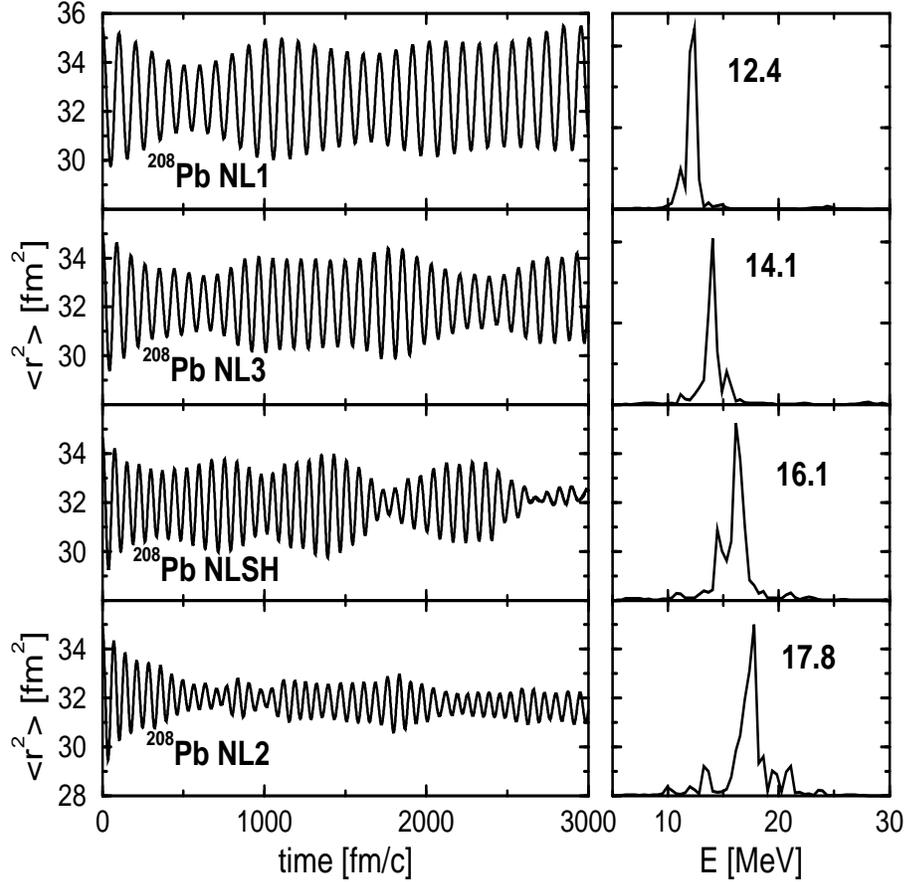
Time-dependent isoscalar monopole moments
$<r^2>(t)$ and the corresponding Fourier power spectra for
$^{208}$Pb. The parameter sets are NL1, NL3, NL-SH and NL2.
\item{\small {\bf FIG. 5.}}
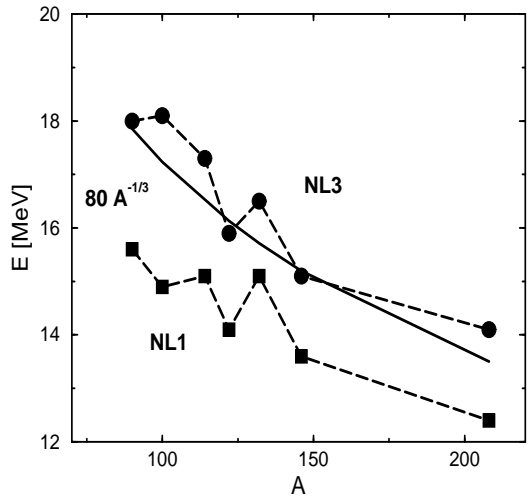
Excitation energies of isoscalar giant
monopole resonances in spherical nuclei as
function of the mass number. The effective interactions
are: NL1 (squares) and NL3 (circles). The solid curve
corresponds to the empirical relation $\approx 80~A^{-1/3}$
MeV.
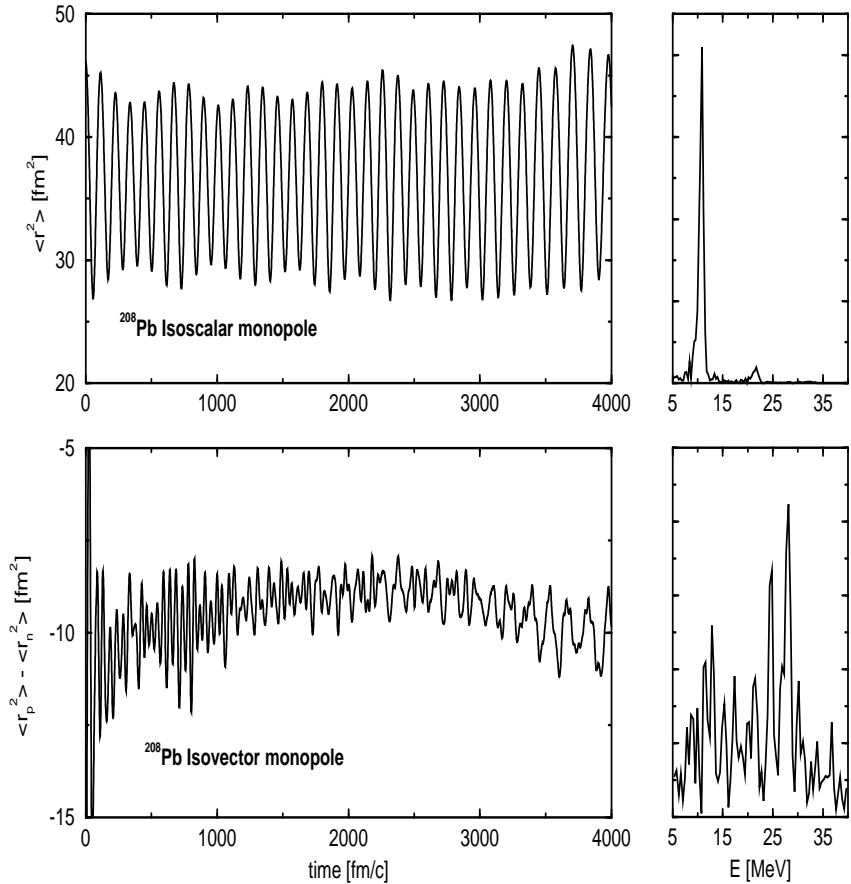
\item{\small {\bf FIG. 6.}}
Results of time-dependent
relativistic mean-field calculations for isoscalar
and isovector oscillations in $^{208}$Pb.
\item{\small {\bf FIG. 7.}}
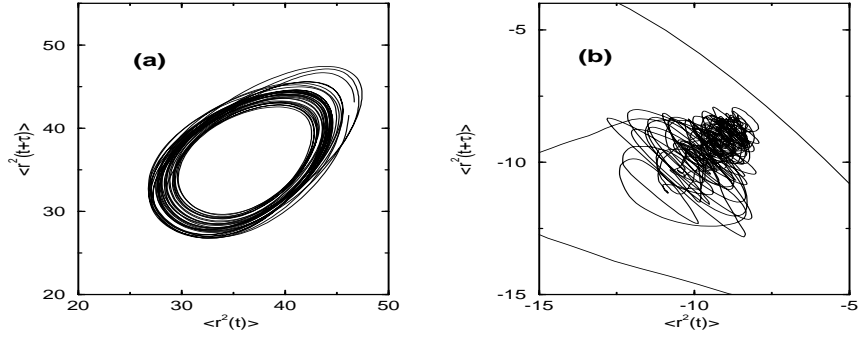
Pseudo-phase space for
isoscalar (a) and isovector (b) monopole oscillations
in $^{208}$Pb.
\item{\small {\bf FIG. 8.}}
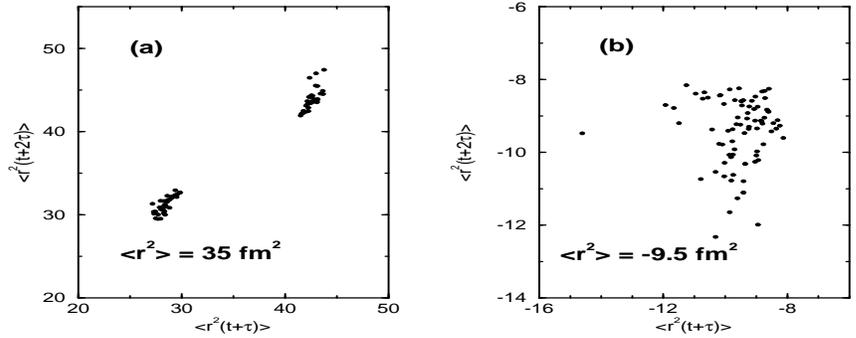
Poincar\' e sections for isoscalar (a)
and isovector (b) monopole oscillations
in $^{208}$Pb.
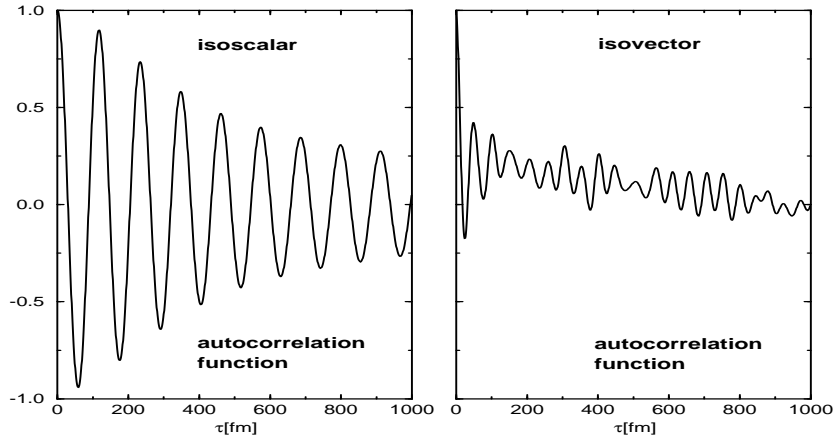
\item{\small {\bf FIG. 9.}}
Autocorrelation functions for isoscalar and isovector oscillations
in $^{208}$Pb.
\end{description}
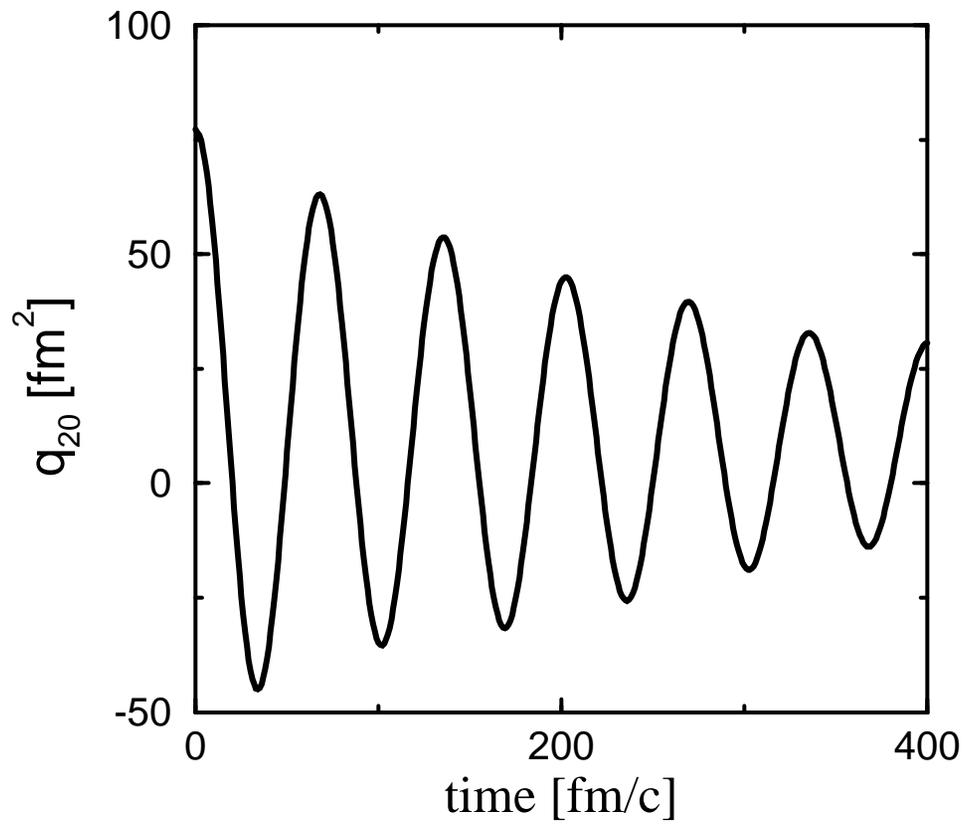
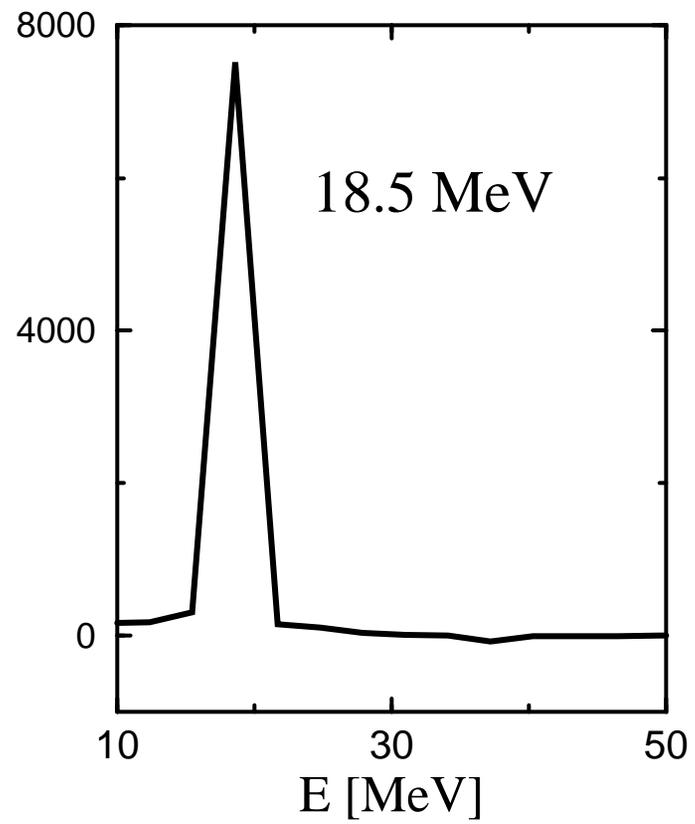
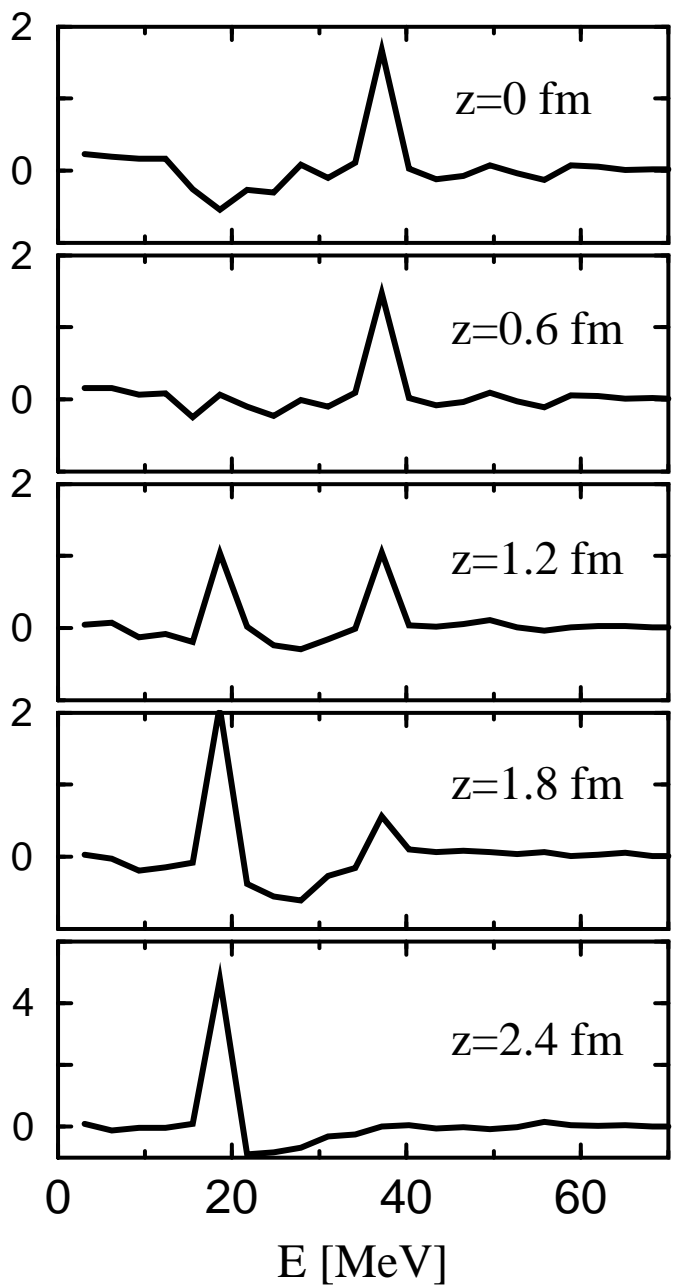
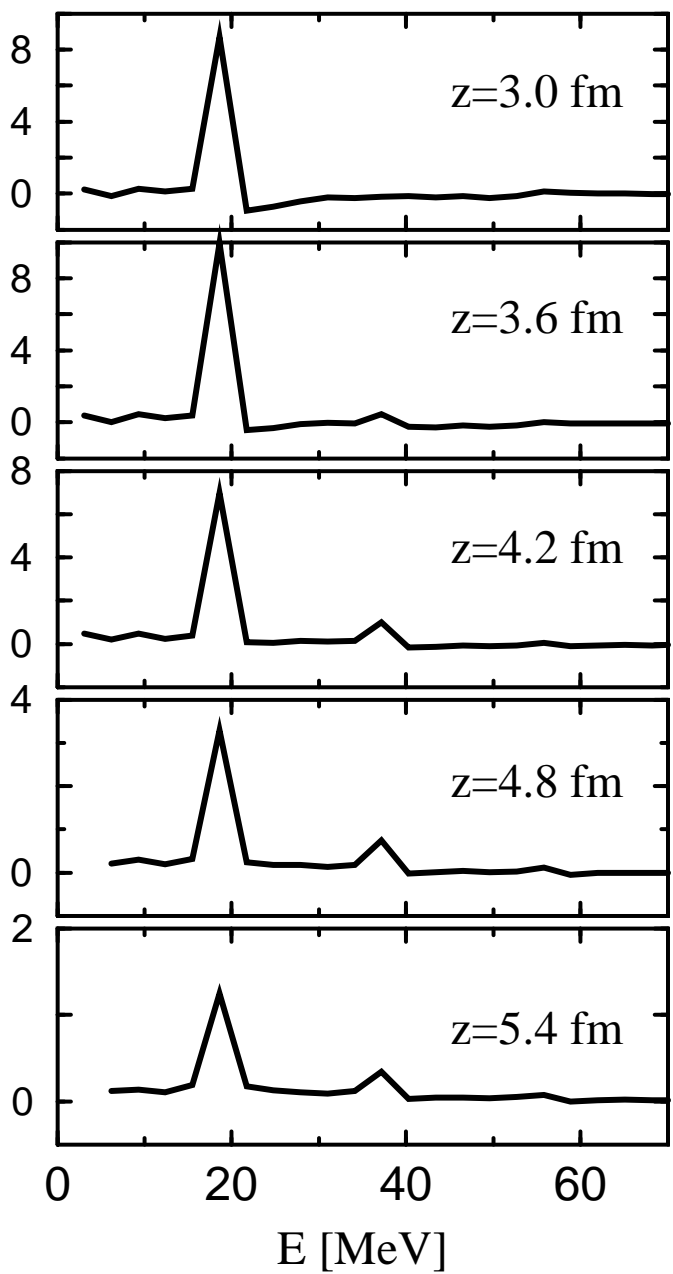
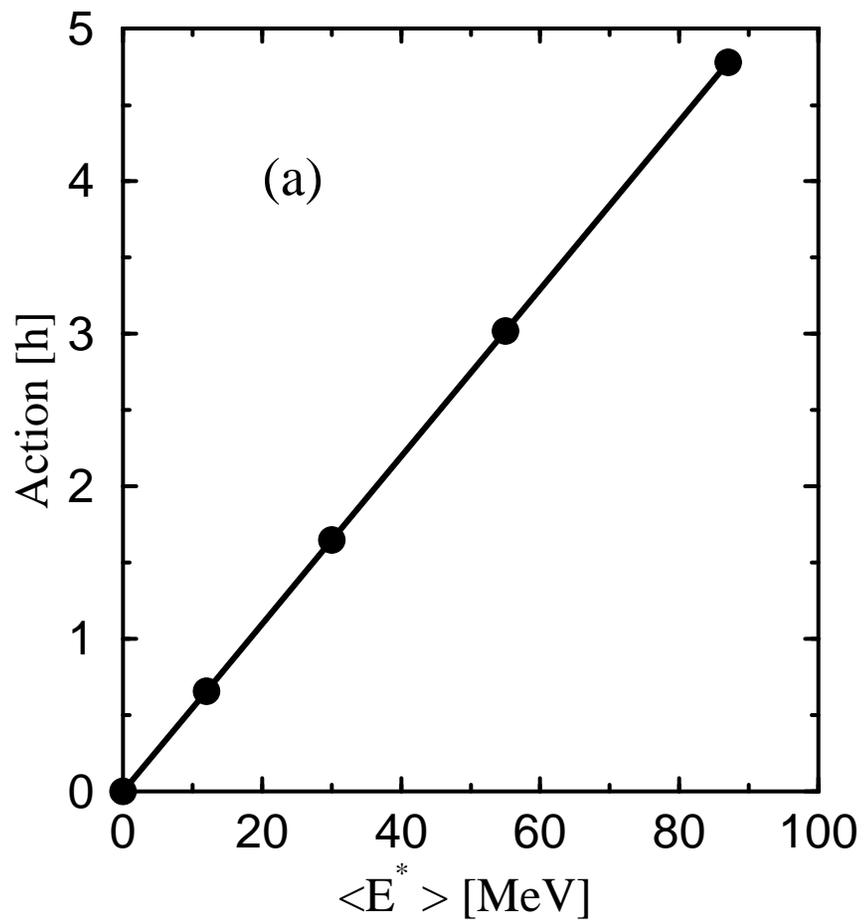
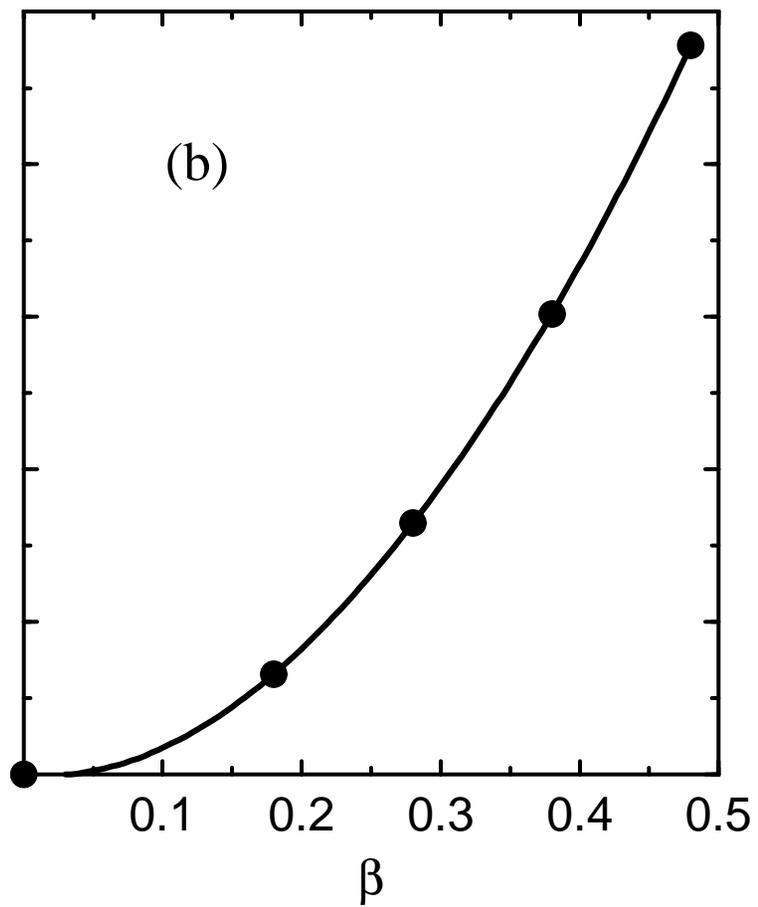
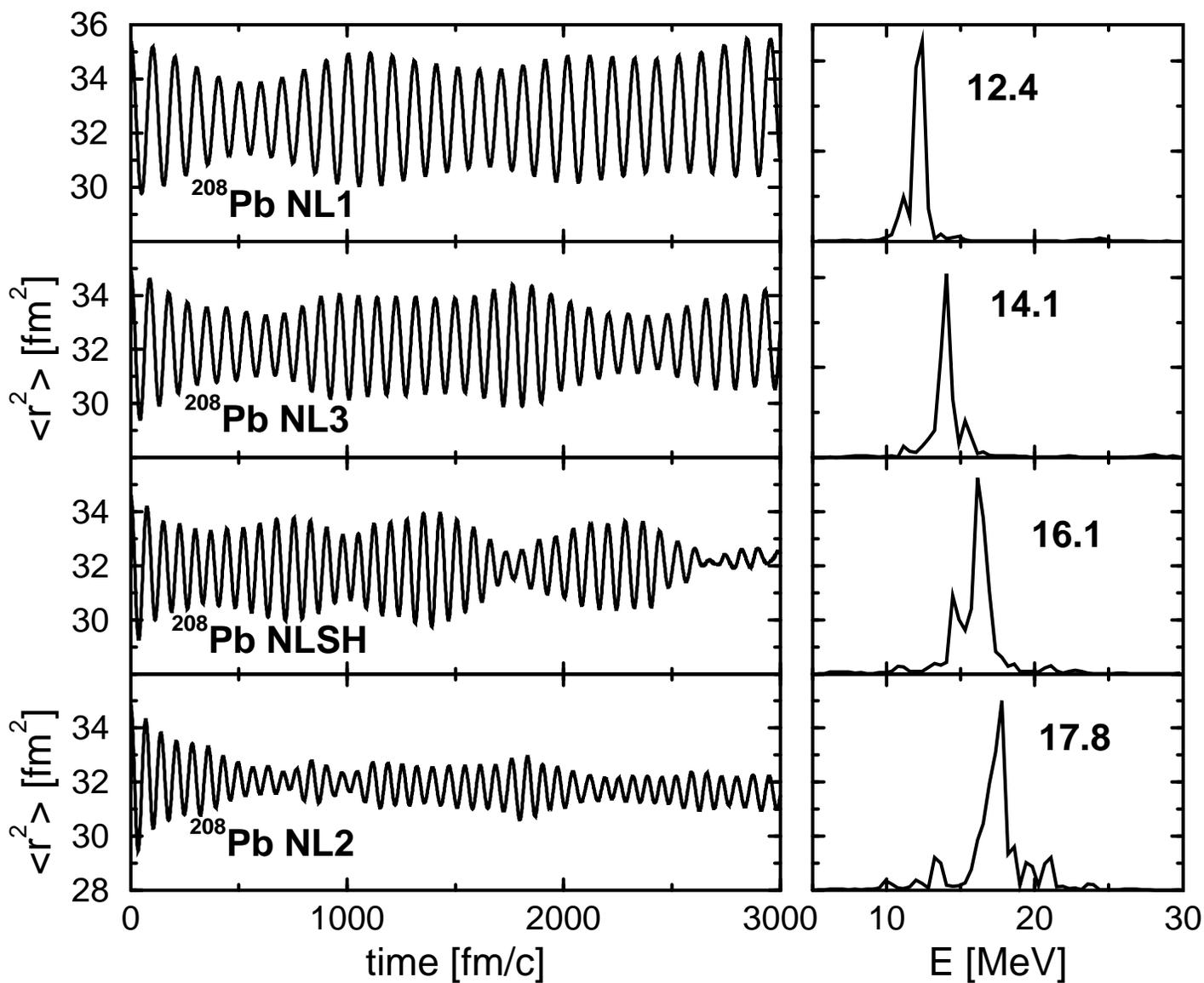
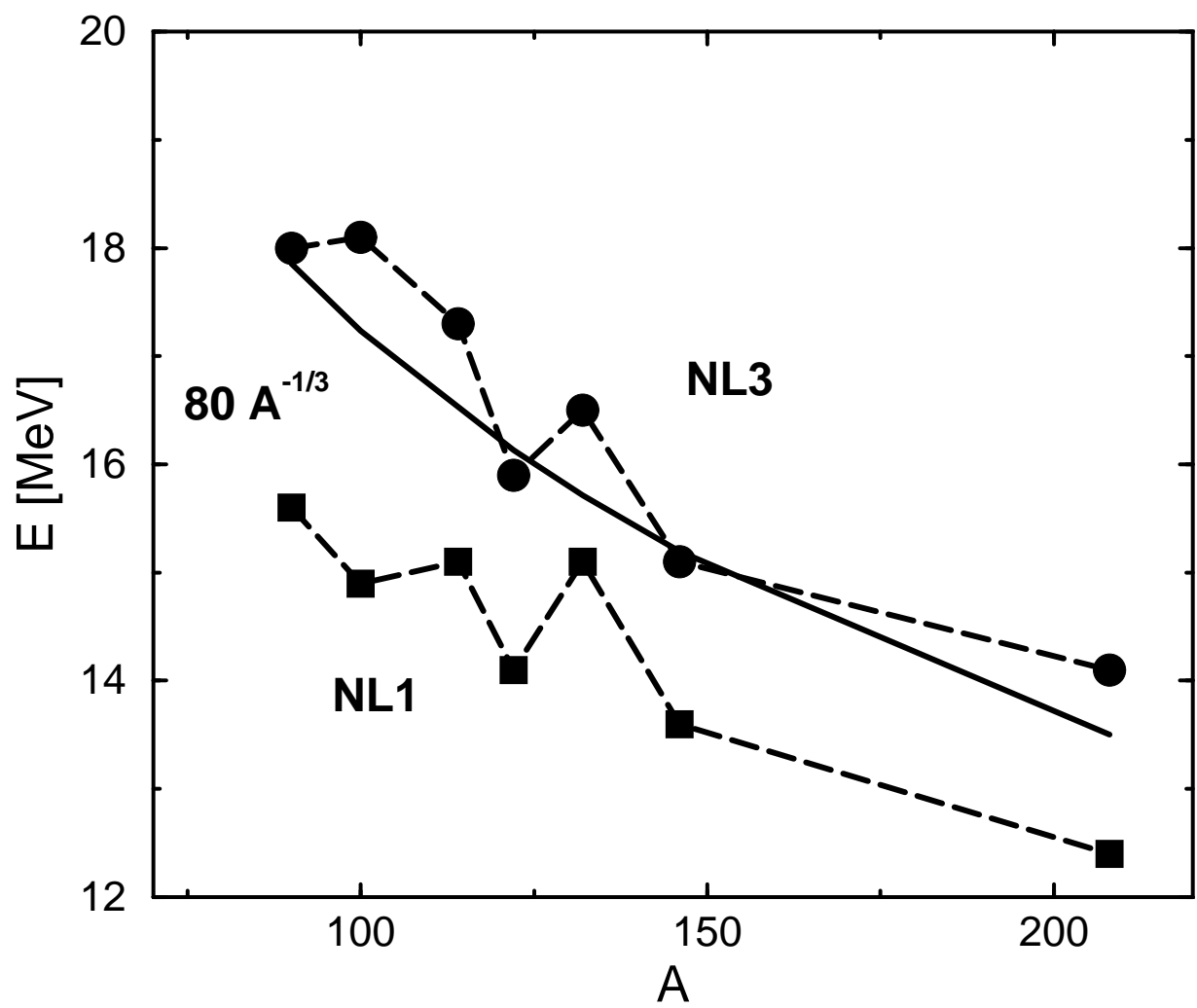
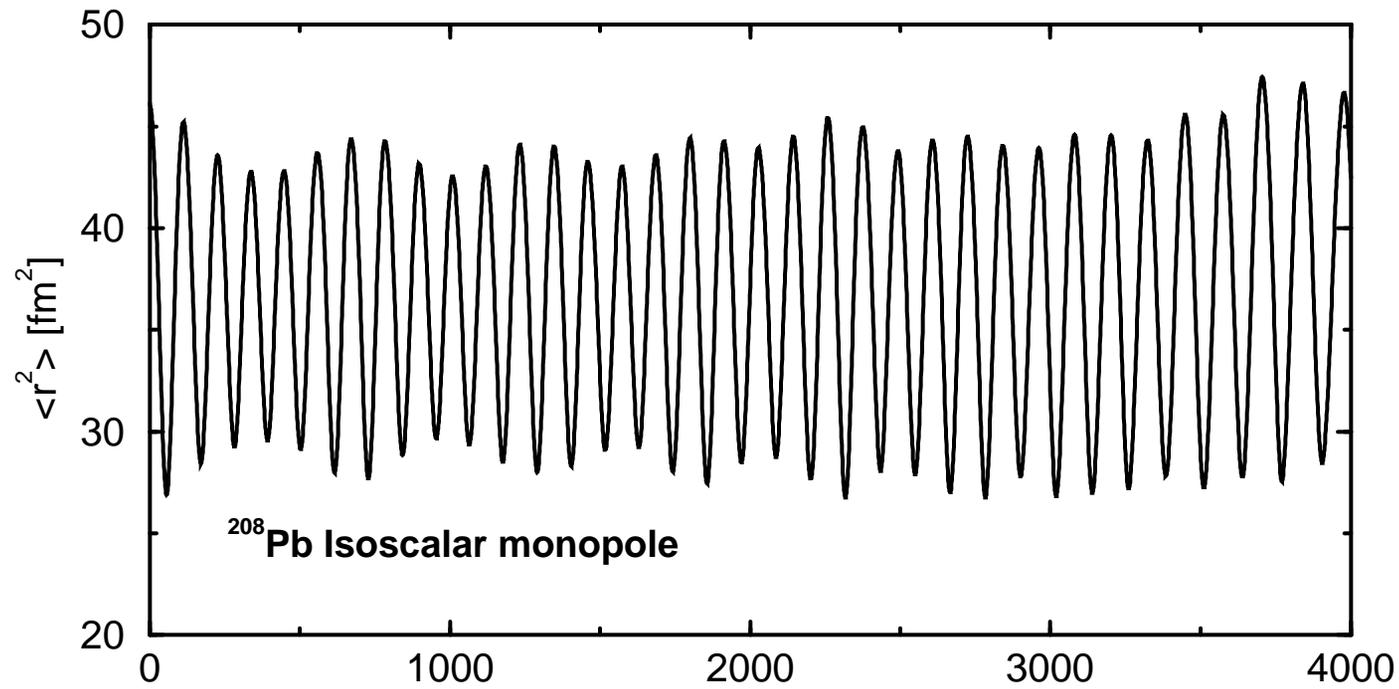
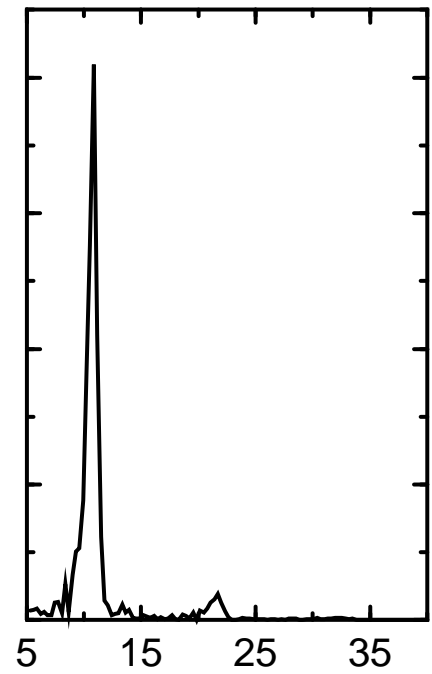
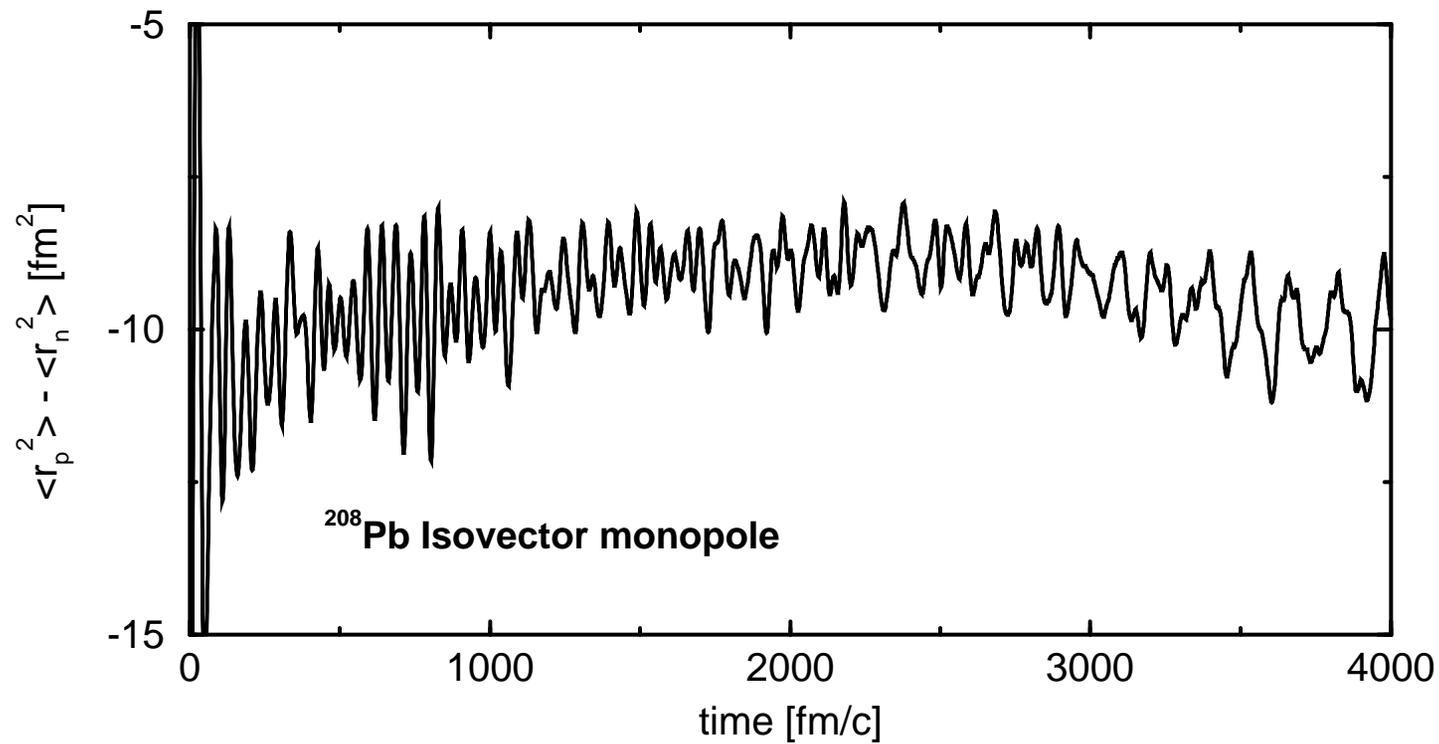
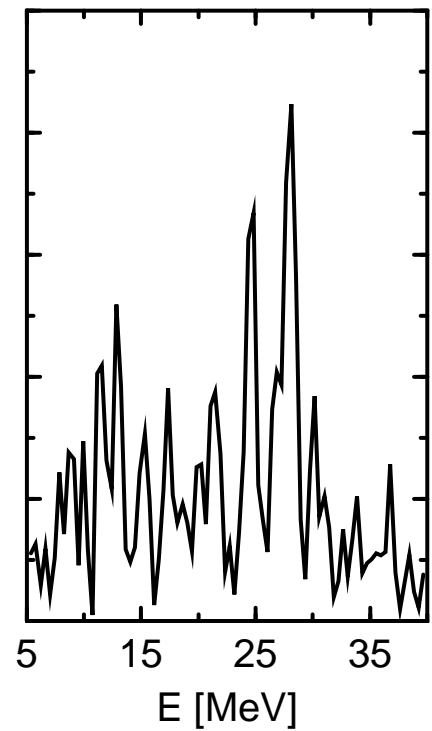
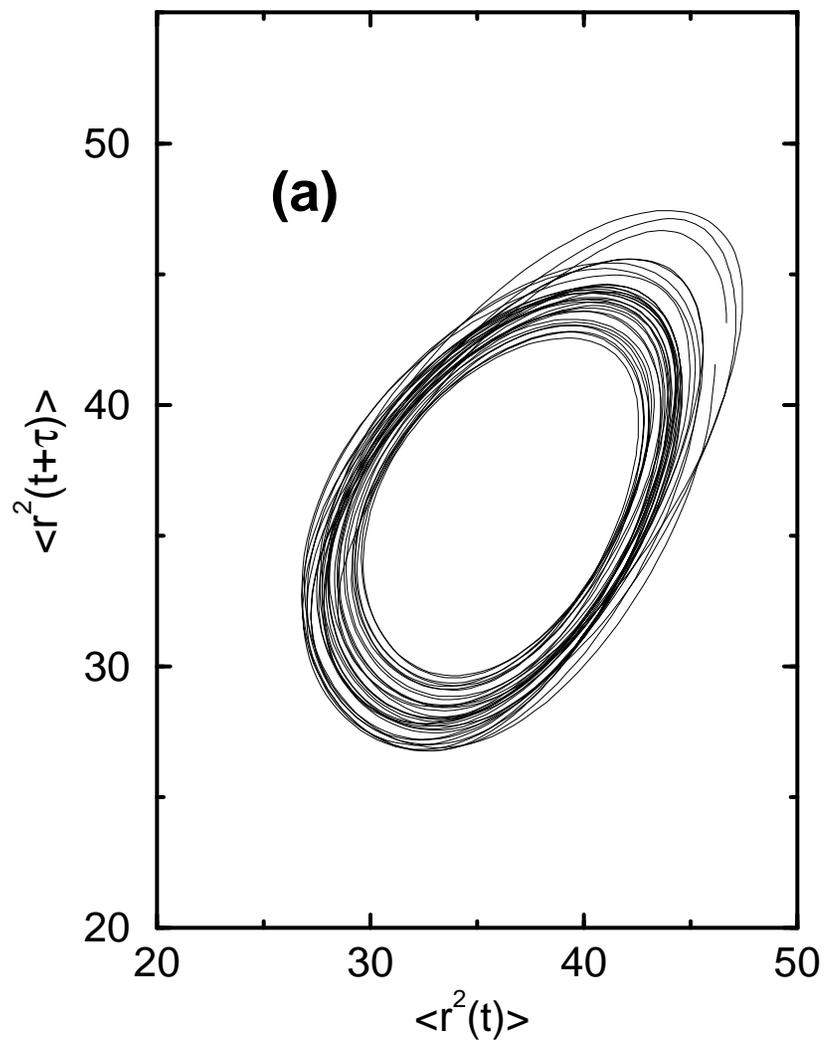
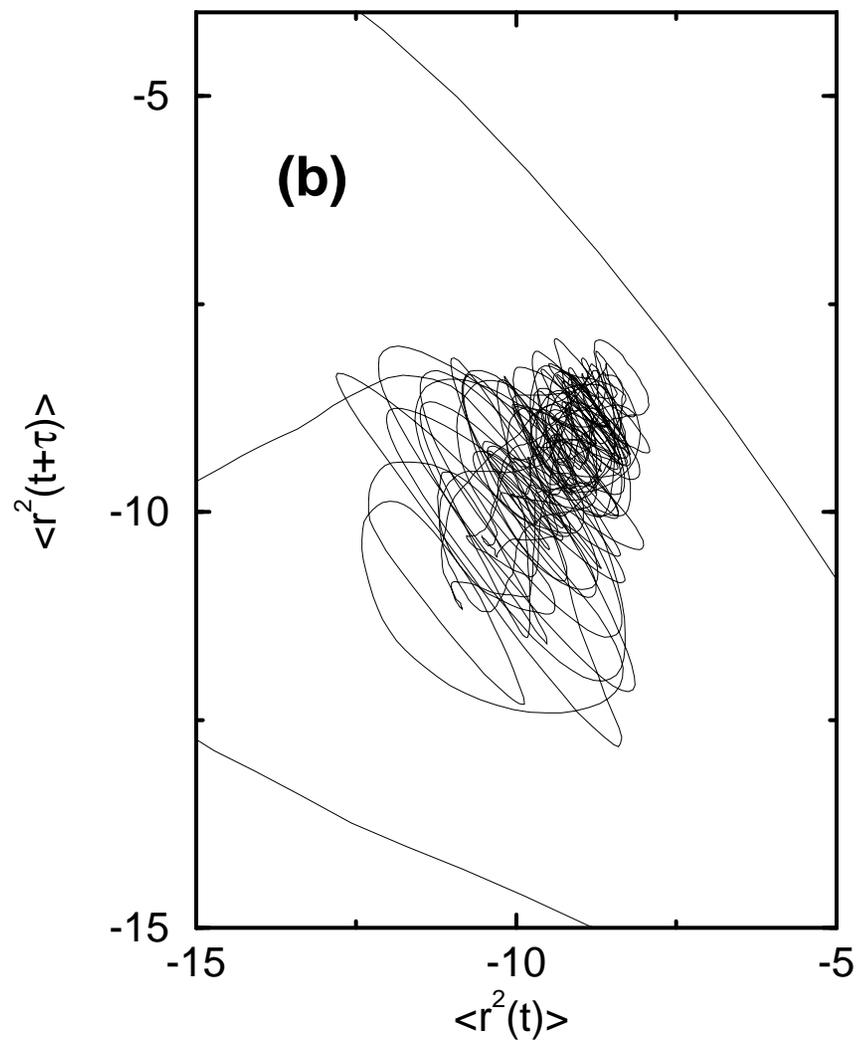
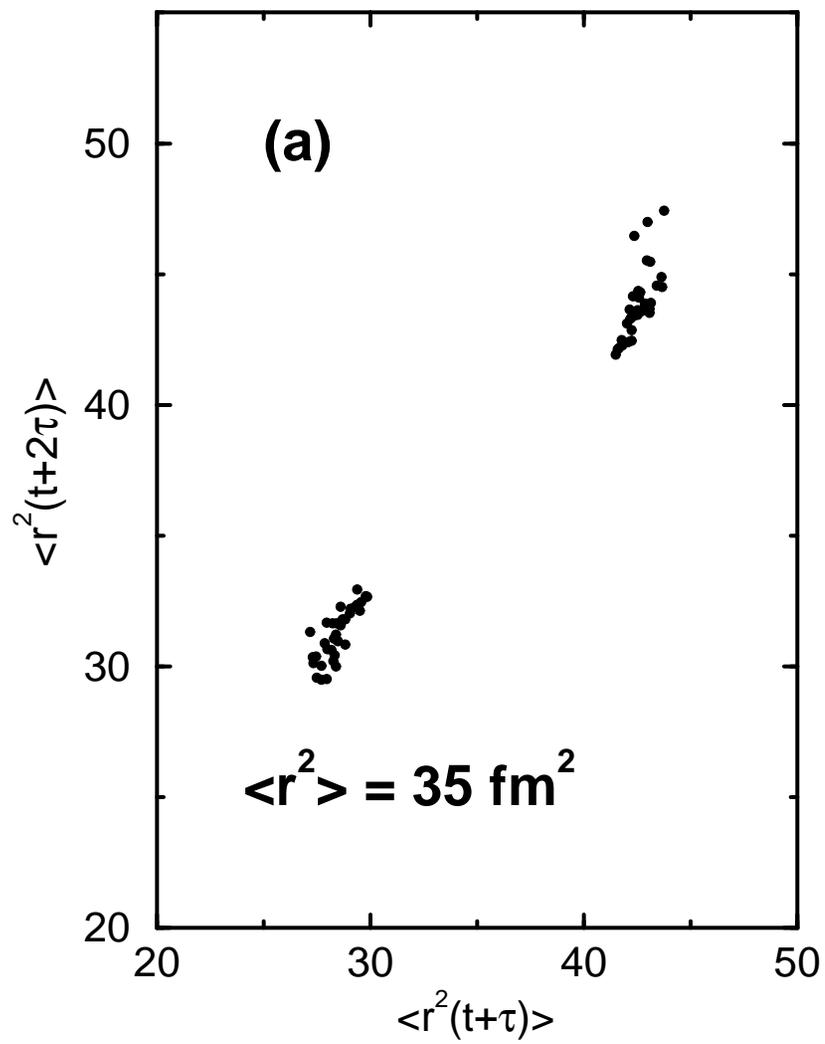
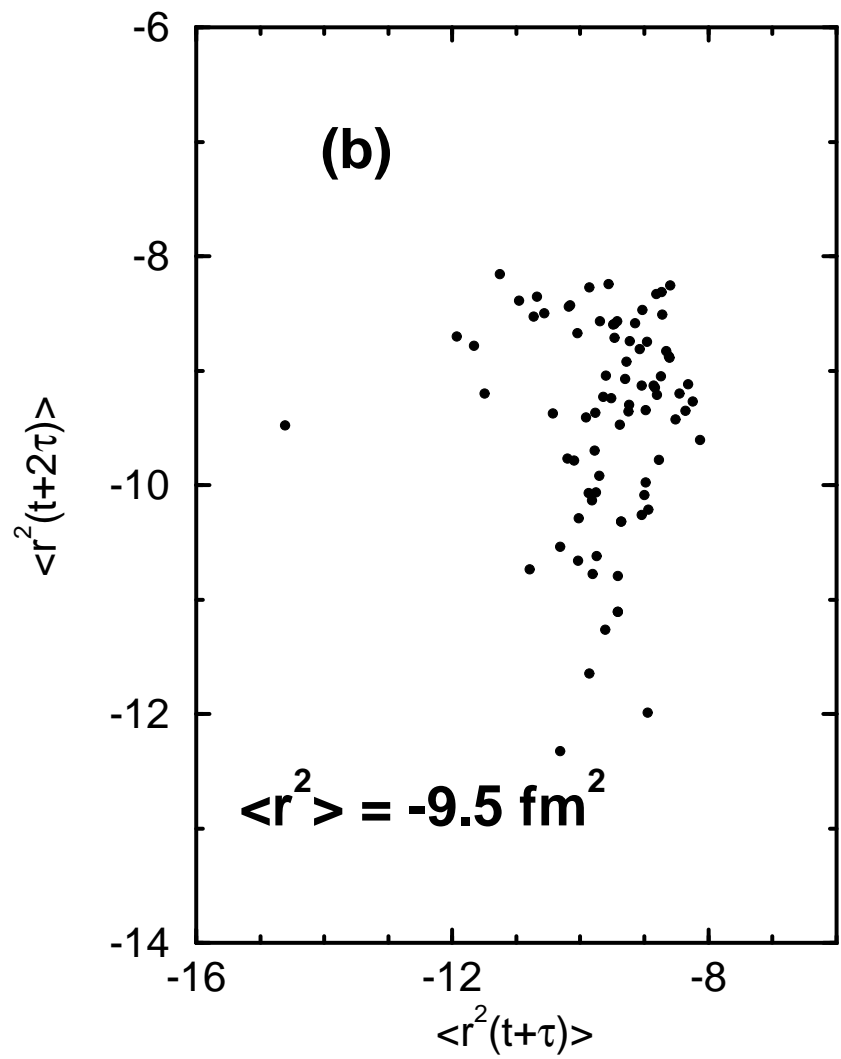
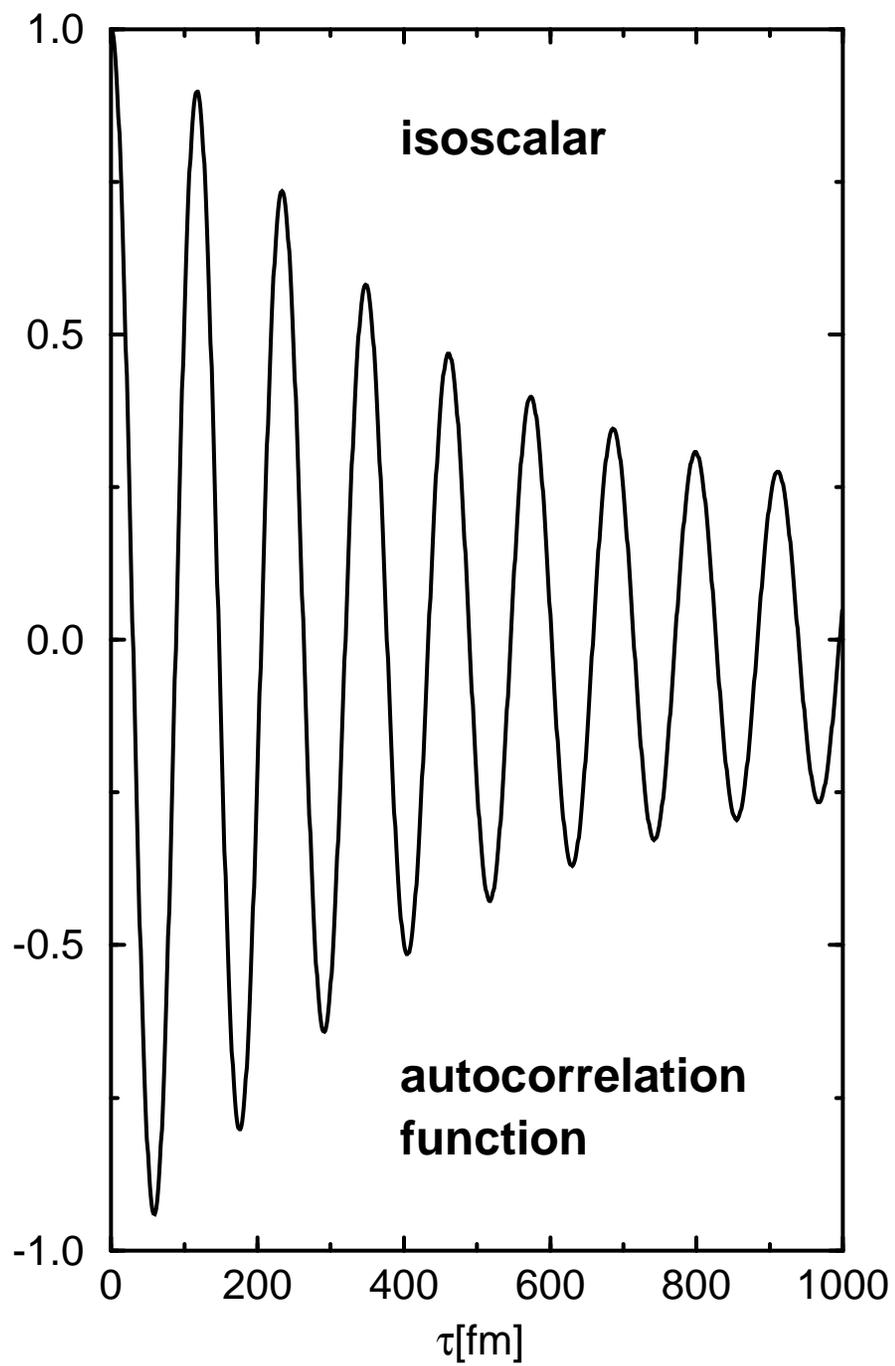
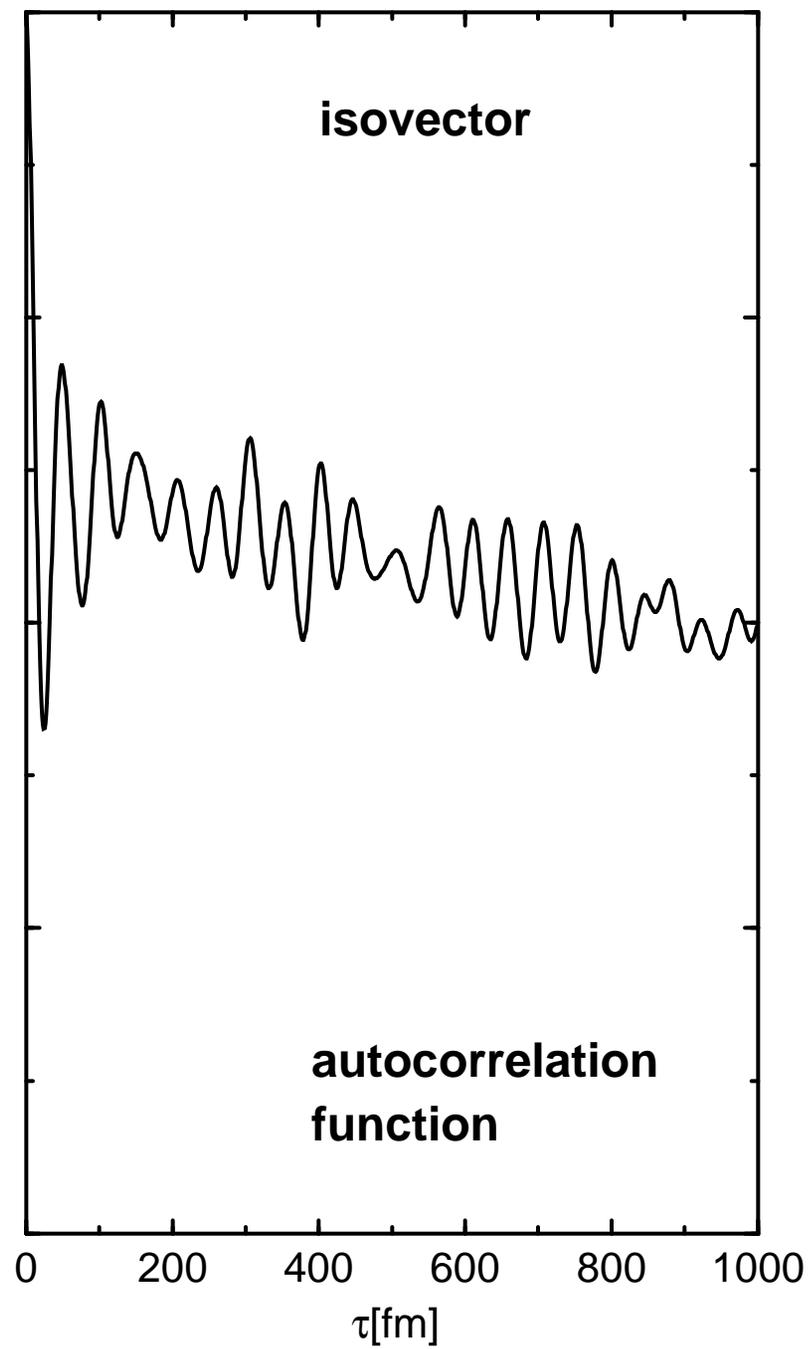
\end{document}